\documentclass[11pt,a4paper,reprint,notitlepage,superscriptaddress,aps,pra]{revtex4-1}

\usepackage{amsmath,amssymb,amsfonts} 

\usepackage{microtype}

\usepackage[svgnames,dvipsnames]{xcolor}
\usepackage{graphicx}
\definecolor{NewBlue}{rgb}{0.1, 0.1, 0.7}
\definecolor{NewRed}{rgb}{0.7, 0.1, 0.1}
\usepackage[colorlinks,
    linkcolor=NewRed,
    citecolor=NewBlue,
    urlcolor=NewRed]{hyperref}

\usepackage[capitalise]{cleveref}

\usepackage[
    exponent-product=\cdot,
    range-phrase=\text{--},
    range-units=single,
    separate-uncertainty,
]{siunitx}

\renewcommand{\t}[1]{\mathrm{{#1}}}

\newcommand{\avg}[1]{\langle {#1}\rangle}

\newcommand{\e}{\mathrm{e}}
\newcommand{\dd}{\mathrm{d}}
\newcommand{\ii}{\mathrm{i}}

\newcommand{\Nph}{10.8}
\newcommand{\dNph}{0.8}
\newcommand{\Tfinal}{77} 
\newcommand{\fzero}{0.43} 
\newcommand{\feff}{148} 
\newcommand{\tdelay}{0.9} 
\newcommand{\Parm}{200} 
\newcommand{\sqz}{3} 
\newcommand{\nimp}{3.5\cdot 10^{-13}}
\newcommand{\nba}{\num{1.0e12}}

\newcommand{\LigoMIT}{LIGO, Massachusetts Institute of Technology, Cambridge, MA 02139, USA}
\newcommand{\LHO}{LIGO Hanford Observatory, Richland, WA 99352, USA}

\begin{document}

\title{Approaching the motional ground state of a 10 kg object}

\author{Chris Whittle}
\affiliation{\LigoMIT}
\author{Evan D. Hall}
\affiliation{\LigoMIT}
\author{Sheila Dwyer}
\affiliation{\LHO}
\author{Nergis Mavalvala}
\affiliation{\LigoMIT}
\author{Vivishek Sudhir}
\affiliation{\LigoMIT}
\affiliation{Department of Mechanical Engineering, Massachusetts Institute of Technology, Cambridge, MA 02139}
\email{vivishek@mit.edu}


\author{R.~Abbott}
\affiliation{LIGO, California Institute of Technology, Pasadena, CA 91125, USA}
\author{A.~Ananyeva}
\affiliation{LIGO, California Institute of Technology, Pasadena, CA 91125, USA}
\author{C.~Austin}
\affiliation{Louisiana State University, Baton Rouge, LA 70803, USA}
\author{L.~Barsotti}
\affiliation{\LigoMIT}
\author{J.~Betzwieser}
\affiliation{LIGO Livingston Observatory, Livingston, LA 70754, USA}
% ...





\begin{abstract}
The motion of a mechanical object\,---\,even a human-sized object\,---\,should be
governed by the rules of quantum mechanics. 
Coaxing them into a quantum state is, however, difficult:
the thermal environment masks any quantum signature of the object's motion. Indeed, the thermal environment also
masks effects of proposed modifications of quantum mechanics at large mass scales.
We prepare the center-of-mass motion of a $\SI{10}{kg}$ mechanical oscillator
in a state with an average phonon occupation of $\Nph$.
The reduction in temperature, from room temperature to \SI{\Tfinal}{\nano\kelvin}, 
is commensurate with an 11 orders-of-magnitude suppression of
quantum back-action by feedback\,---\,and a 13 
orders-of-magnitude increase in the mass of an object prepared
close to its motional ground state. This begets the possibility of probing gravity on 
massive quantum systems.
\end{abstract}

\maketitle

The apparent classical behavior of tangibly massive objects is, according to
conventional quantum mechanics, the symptom of decoherence. 
Thermal decoherence, caused by the interaction of a system with a thermal environment, 
is by far the most pervasive.
For a mechanical
oscillator of mass $m$ and natural frequency $\Omega_0$, 
thermal decoherence induces motion characterized by
spectral density $S_x^\t{th}[\Omega_0] =
(2n_\t{th}[\Omega_0]+1)S_x^\t{zp}[\Omega_0]$, where $n_\t{th}[\Omega_0]\approx
k_B T/\hbar \Omega_0$ is the average thermal phonon occupation due to the
environment (at temperature $T$) and $S_x^\t{zp}[\Omega_0] = 8x_\t{zp}^2/\Gamma_0[\Omega_0]$ is
its motional zero-point fluctuation, $x_\t{zp}=\sqrt{\hbar/(2m\Omega_0)}$,
concentrated in a frequency band of width $\Gamma_0[\Omega_0]$.
Thermal fluctuations obscure signatures of decoherence that allegedly 
arise from modifications of quantum mechanics at large masses \cite{Kar66,Dio89,Pen96,Bassi17},
and limit the sensitivity of mechanical transducers in metrology applications \cite{Saul90}.
Techniques to probe both frontiers
call for large mass mechanical objects prepared in pure quantum states.

Over the past decade, progressively larger objects all the way to 
nanomechanical oscillators have been prepared in their motional ground state
\cite{ChanPain11,Teuf11,PetReg16,RosSch18,DelAsp20}.  A vast majority of these
experiments rely on isolating the oscillator in an elastic or electromagnetic
trap in the $\gtrsim \SI{100}{\kHz}$ frequency range, embedded in a
sideband-resolved electromagnetic cavity, typically in a cryogenic environment.
These methods do not address a number of technical challenges
unique to mechanical oscillators above the milli-/gram mass scale.
For one, the large optical power required to trap massive oscillators introduces
extraneous heating and other opto-mechanical nonlinearities.
Meanwhile, the low resonant frequency of large suspended oscillators doubly compounds 
the problem of thermal decoherence by increasing the intrinsic thermal motion
($n_\t{tot}\propto 1/\Omega_0$) and precluding efficient cavity sideband
cooling.  
Therefore a different route is needed to prepare large-mass oscillators in pure quantum states.

\begin{figure*}[t!]
	\centering
	\includegraphics[width=0.9\textwidth]{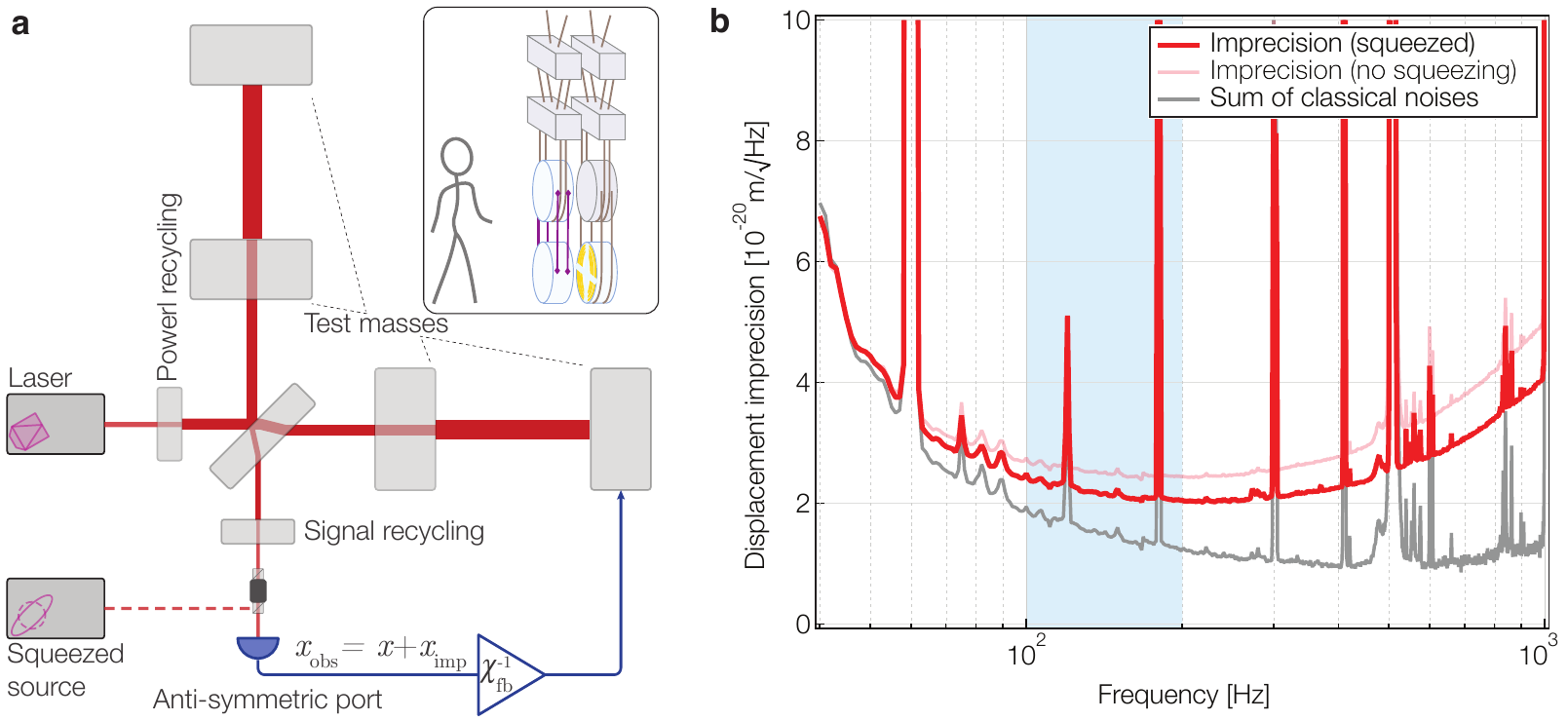}
	\caption{\label{fig:ligo}
	Advanced LIGO interferometer.
	(A) Laser light is split and recombined at
	a beam-splitter, forming a Michelson interferometer.  
    Its response is shaped by the Fabry--P\'{e}rot cavities in the arms and the signal-recycling mirror.
    The power-recycling mirror and injection of squeezed light enhances the sensitivity. 
	Inset shows the suspension system of each of the four \SI{40}{\kg} mirrors:
	the final mass on the forward chain is the
	\SI{40}{\kg} mirror, suspended on fused silica wires (purple) featuring a quality factor $Q\approx 8\cdot 10^7$; they
	can be displaced by electrostatic forces due to voltages applied on electrodes
	(yellow) etched onto the reaction mass suspended behind it; average human sketched for scale.  
	(B) The displacement sensitivity (red) of the interferometer is
	\SI{2e-20}{\meter/\sqrt{\Hz}} at \SIrange{100}{200}{\Hz}, where it is largely
	shot-noise (light red), suppressed by about \SI{\sqz}{dB} from injection of squeezed
	vacuum (red), and a combination of extraneous technical noises (gray). 
	Blue band shows the frequency interval in which the pendulum mode is trapped
	and cooled.
}
\end{figure*}

The Advanced LIGO gravitational-wave detectors offer a unique perspective on this problem. 
Advanced LIGO is a pair of Michelson interferometers, each with \SI{4}{\km} long Fabry--P\'{e}rot arm cavities formed
by \SI{40}{\kg} mirrors that hang on fused silica fibers (\cref{fig:ligo}).
The differential motion of each pair of arm cavity mirrors forms a mechanical
oscillator with a reduced mass of \SI{20}{\kg}; the differential motion of each
such oscillator in either arm, sensed by the Michelson interferometer, forms a
mechanical oscillator of effective mass $m=\SI{10}{\kilo\gram}$ that is the object of
our attention.  The oscillator follows the pendulum-like motion of the
suspended mirror at a frequency $\Omega_0 \approx 2\pi\cdot \SI{\fzero}{\Hz}$;
gravitational stress dilution is expected to realize a quality
factor of $Q_0 \approx 10^8$ \cite{Cumm12}.
Its displacement fluctuates due to the presence of $n_\t{th}[\Omega_0]\approx 10^{13}$ phonons.
The interferometer resonantly transduces the differential arm motion into
optical power fluctuations at the anti-symmetric
port, which is sensed by homodyne detection; 
during ordinary operation,
these fluctuations encode the gravitational-wave signals.
In this state, the homodyne photocurrent
fluctuations bear the apparent displacement $\delta x_\t{obs} = \delta x
+\delta x_\t{imp}$;
here $\delta x$ is the physical motion of the differential arm, which contains
the displacement of the oscillator, and $\delta x_\t{imp}$ is the measurement
imprecision. 
The imprecision noise, depicted in \cref{fig:ligo}B,
is \SI{2e-20}{\meter/\sqrt{\Hz}} around \SIrange{100}{200}{\Hz} and is largely
quantum shot noise\,---\,suppressed by $\sim \SI{\sqz}{dB}$ by injection of squeezed
light~\cite{Tse19}, and shaped by the response of the signal recycling cavity
--- with a secondary contribution from mechanical dissipation in the mirror
coatings \cite{O3_performance}.  This sensitivity is equivalent to $n_\t{imp}\equiv
S_x^\t{imp}/2S_x^\t{zp}\approx \nimp$ phonons for a $10\, \t{kg}$
oscillator at $\sim \SI{150}{\Hz}$\,---\,a record low number
(Ref.~\cite{RosSch18} demonstrates $n_\t{imp}\approx 10^{-7}$) tantamount to
resolving the zero-point motion of the oscillator with $\sim 125\, \t{dB}$
signal to squeezed-shot-noise ratio, and comparable to the requirement to
feedback cool the oscillator to its ground state ($n_\t{imp}\sim 1/2n_\t{th}$,
for a viscously-damped oscillator \cite{WilKip15}).

In order to take advantage of this precision, we actively
stiffen the pendulum mode by synthesizing a force proportional to the observed
displacement (i.e. $\propto \Omega_\t{fb}^2 \delta x_\t{obs}$) and in-phase
with the motion $\delta x$, trapping the pendulum mode as an oscillator around
$\Omega_\t{fb}\approx 2\pi \cdot \SI{\feff}{\Hz}$.  Two additional sources of
decoherence plague this scheme. First, such measurement
precision comes at the expense of additional quantum back-action on the
pendulum mode: radiation pressure shot noise from the \SI{\Parm}{\kW} intracavity power and
the anti-squeezed intracavity field produces motion \cite{hao20} equivalent
to about $n_\t{ba}[\Omega_0]\approx \nba$ phonons.
However, as long as the measurement record resolves the quantum
back-action at a rate comparable to the thermal decoherence, active feedback
can suppress it \cite{WilKip15,SudKip17,RosSch18}.  Secondly, the
feedback of amplified imprecision noise leads to an additional ``feedback
back-action'', $n_\t{fb} \approx Q_0^2(\Omega_\t{fb}/\Omega_0)^4 n_\t{imp}$ (see Supplementary Information), 
which increases with the trap frequency.
However, this is partially compensated by the $\Omega_\t{fb}/\Omega_0 \approx 300$ fold reduction in 
both the thermal occupation and decay rate of the trapped oscillator due to structural damping \cite{Saul90}.

\begin{figure*}[t!]
    \centering
    \includegraphics[width=\textwidth]{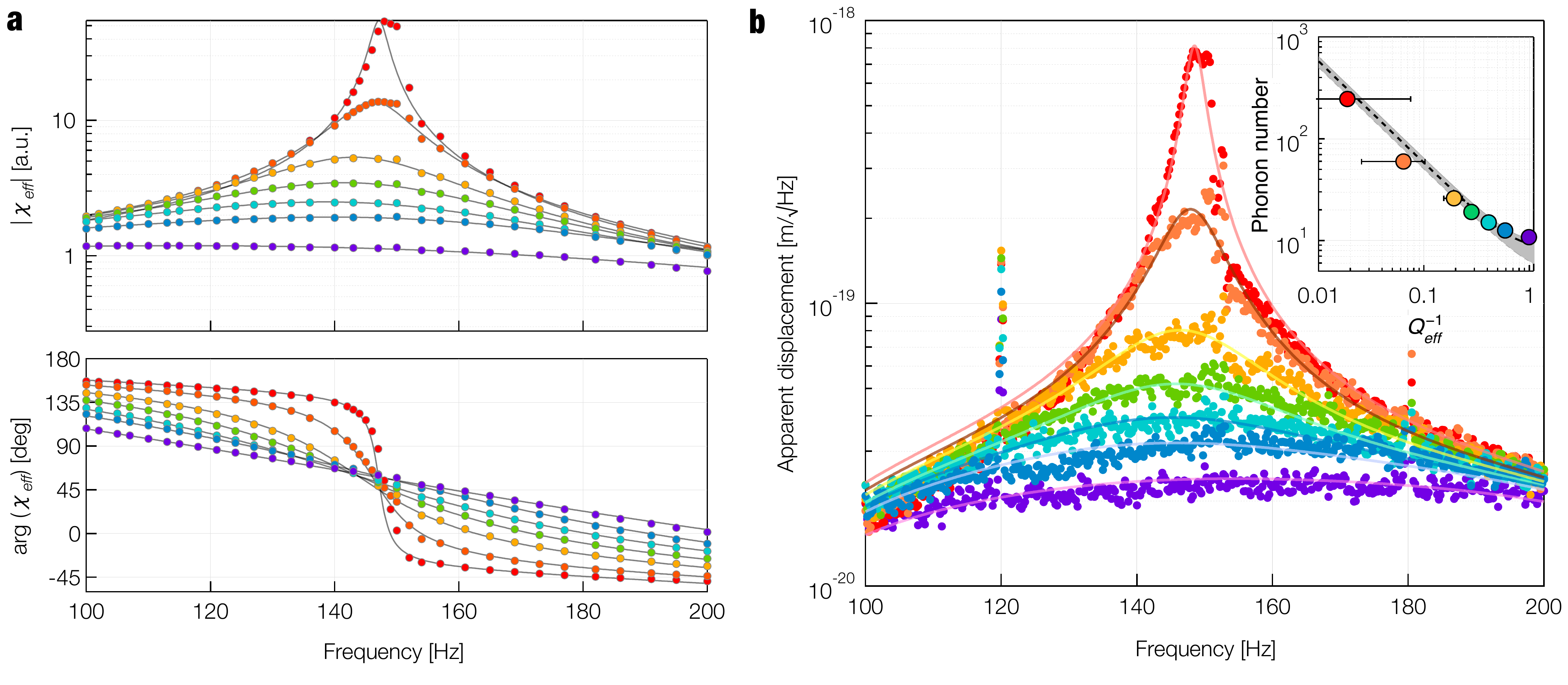}
    \caption{\label{fig2}
        Trapping and cooling of a 10 kg oscillator to 10 quanta.
        (A) Effective susceptibility of the oscillator for each setting 
        of the damping filter, measured by exciting the feedback loop
        at each frequency and demodulating its response at the same frequency.
        The lines show fits to a model of the susceptibility of a damped harmonic oscillator
        with an additional delay, i.e. $\chi_\t{eff}[\Omega]e^{i\Omega\tau}$; fits to
        the phase response produce $\tau = \SI{\tdelay}{\milli\second}$.
        (B) Displacement spectrum of the oscillator as the damping is increased. Solid
        lines show fits to a model of the observed spectrum $S_x^\t{obs}$ (see text for details)
        where the effective susceptibility is determined by the response measurements in panel (A),
        and only the frequency-dependent imprecision noise and force noise are variable.
        Inset shows the inferred average phonon occupation for each of the curves in the main panel,
        as a function of the damping quality factor;
        also shown is a model (black dashed) with model uncertainties (gray band).
        (The disagreement between the simple model and data --- both the transfer functions and spectra --- 
        around \SIrange{150}{155}{\hertz} arises from a coupling between the motion of the 
        pendulum and the upper intermediate mass of the suspension~\cite{Sun20}.)
    }
\end{figure*}

To trap and damp the oscillator, we adjust the feedback control so that $\delta
F_\text{fb} = \chi_\t{fb}^{-1}\delta x_\t{obs}$, with a feedback filter,
$\chi_\t{fb}^{-1}\propto \Omega_\t{fb}^2 + \ii \Omega \Gamma_\text{fb}$,
between \SIrange{100}{200}{\Hz}.  This is implemented by careful shaping of the
control loop that is otherwise used to stabilize the interferometer at its
linear operating point.  
The feedback force is applied on the mirror electrostatically: 
gold electrodes on the reaction mass (\cref{fig:ligo}A) are held at 
a $\SI{400}{\volt}$ bias, whose fringing field polarizes the dielectric test mass; control voltages 
added on interleaved electrodes produce a proportional force (extraneous force noise 
produces $\ll 1$ phonon of excess occupation on average, see Supplementary Information).
The overall feedback gain is adjusted so that the
system's effective susceptibility takes the form, $\chi_\text{eff}[\Omega]
\propto (-\Omega^2 +\Omega_\t{eff}^2 +\ii \Omega
\Gamma_\t{eff}[\Omega])^{-1}/m$, of that of an oscillator with frequency
$\Omega_\t{eff} = \sqrt{\Omega_0^2 + \Omega_\t{fb}^2} \approx \Omega_\t{fb}
\approx 2\pi\cdot\SI{\feff}{\Hz}$.
Delays in the feedback loop limit the trap frequency and cause the oscillator
to be intrinsically ``cold-damped''. 
In particular, the phase response of the
notch filters used to prevent excitation of the violin modes of the
suspension (at \SI{500}{\hertz} and harmonics, featuring quality factors $\gtrsim
10^9$) in conjunction with the feedback filter leaves the interferometer's length control
system with a phase margin of $1^\circ$ for a trap frequency of $\SI{148}{\hertz}$.
Physical delay in the loop also cold-damps the trapped oscillator to a quality factor
of $\approx 50$ (\cref{fig2}B red trace; see SI for further details).
The oscillator is damped
further by modifying the imaginary part of the feedback filter. \cref{fig2}A
shows the effective susceptibilities of the trapped and damped oscillator so
realized. 
The largest damping rate, corresponding to a quality factor of
$\approx 1$, is limited by the gain margin ($\approx 10^{-3}$) of the
control loop.
Around the trap frequency (\SIrange{100}{200}{\Hz}), additional
force noise on the oscillator due to feedback is dominated by sub-quantum
fluctuations of the squeezed imprecision noise.

The calibrated in-loop signal, depicted in \cref{fig2}B, shows the apparent displacement
fluctuations of the trapped and damped oscillator ($\delta x_\t{obs}$). 
This can be understood using a simple model (see Supplementary Information), $\delta x_\t{obs}=\chi_\t{eff}(\delta
F_\t{th}+\delta F_\t{ba}-\chi_\t{fb}^{-1}\delta x_\t{imp})+\delta x_\t{imp}$. It describes
the oscillator\,---\,with intrinsic susceptibility $\chi_0$\,---\, whose displacement
responds via the feedback-modified effective susceptibility $\chi_\t{eff} =
(\chi_0^{-1}+\chi_\t{fb}^{-1})^{-1}$, to three forces:
a frequency-dependent structural thermal force ($\delta F_\t{th}$), a white quantum measurement 
back-action force ($\delta F_\t{ba}$), and an additional force noise ($\propto 
\chi_\t{fb}^{-1} \delta x_\t{imp}$) due to feedback of
imprecision noise through the feedback filter; and riding
on the imprecision noise ($\delta x_\t{imp}$).
The spectra of the observed displacement $S_x^\t{obs}$ predicted by this model
are shown as the solid lines in \cref{fig2}B.  In the model, the effective
susceptibility is fully determined by the response measurements shown in
\cref{fig2}A, independent of the frequency-dependent force noise and imprecision
noise.  The latter, determined self-consistently amongst the
displacement noise in \cref{fig2}B, shows a variation between the
different feedback settings of less than $1\%$, consistent with expected drift
in the Advanced LIGO interferometer over the $\sim 2\,\t{hr}$ timescale over which the
experiment was performed.  
Several sources of uncertainty are accounted for in this process. Calibration
of the displacement spectra contributes $\approx 2\%$ uncertainty
\cite{Sun20}. Uncertainties in the effective susceptibility
$\chi_\t{eff}$\,---\,from fits to \cref{fig2}A\,---\, are at the $1\%$ level, limited
by the $1\,\t{s}$ averaging used per point in measurements of the
response (see Supplementary Information). The dominant uncertainty is in the fits to the displacement spectra
of \cref{fig2}B using the model for $S_x^\t{obs}$: the frequency-dependence of
the imprecision noise and structurally damped thermal force noise produce a
$\approx 5\%$ variation between the various spectra in \cref{fig2}B.

\begin{figure}[t!]
\centering
\includegraphics[width=\columnwidth]{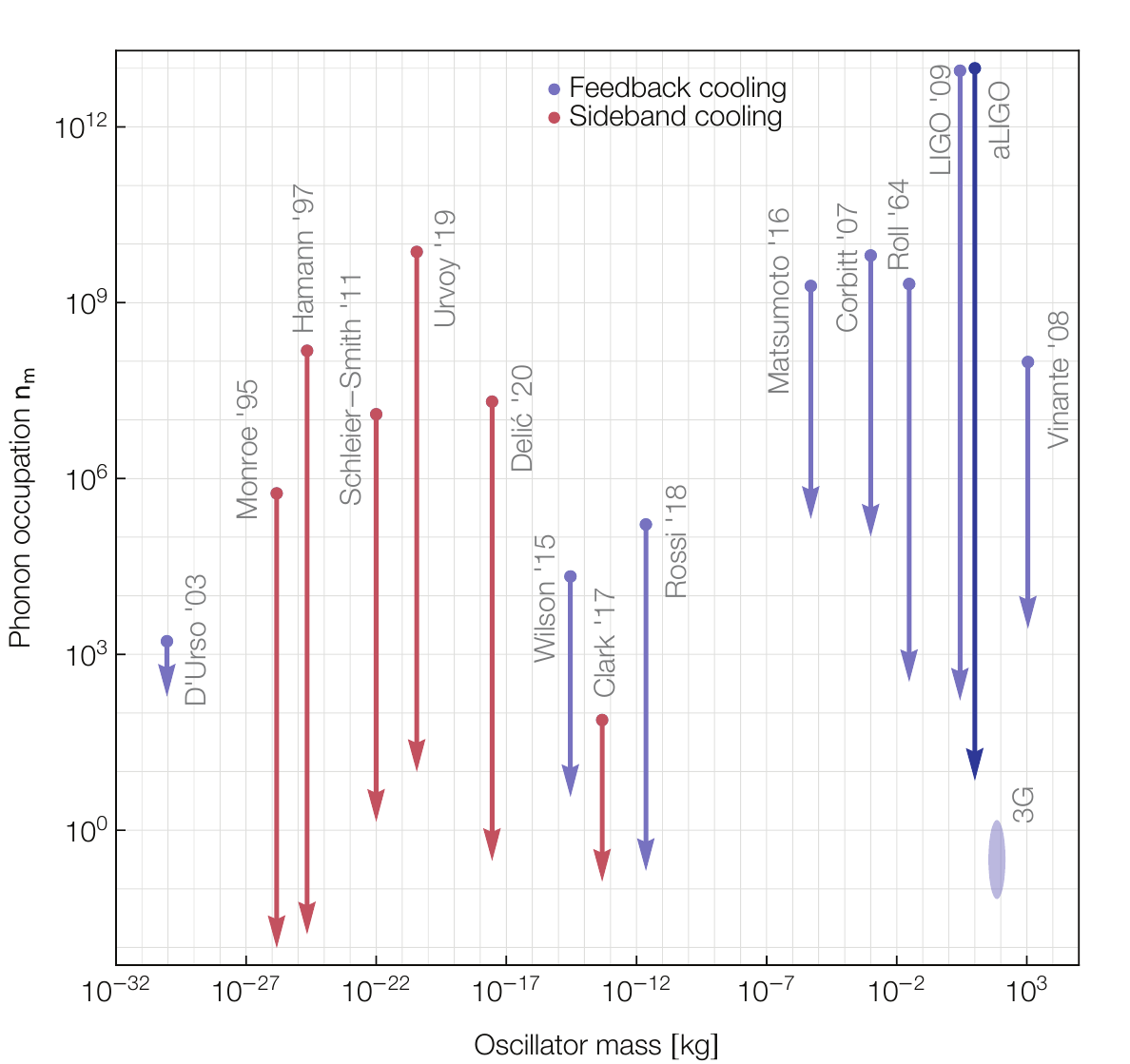}
\caption{\label{fig:review}
A selection of oscillator cooling experiments~\cite{UrsGab03,Monroe1995,Hamann1998,SchleierSmith2011,Urvoy2019,DelAsp20,	WilKip15,Clark2017,RosSch18,Matsumoto2016,Corbitt2007,	roll1964equivalence,iligo_cooling,auriga08}.
	The initial occupations mentioned are those of the relevant oscillator mode as defined 
	by the natural trap frequency, at its ambient temperature. For atomic physics experiments,
	this is usually at room temperature in a harmonic electromagnetic trap; whereas for most
	solid-state mechanical oscillators, it is the harmonic mode defined by the Hookean restoring
	force of its elastic suspension, and typically at cryogenic temperatures (the exception is 
	the recent work from Deli\'{c} et al. \cite{DelAsp20} which demonstrated cavity-cooling of an electromagnetically
	trapped nano-particle to its ground state).
	Our result (``aLIGO'') sets a new record in the macroscopic mass range,
	reaching $\Nph\pm\dNph$ phonons. Experiments with future gravitational-wave
	interferometers (``3G'') will achieve occupations below 1.
}
\end{figure}

The effective phonon occupation ($n_\t{eff}$) of the cooled 
oscillator can be defined through,
$\hbar \Omega_\t{eff}(n_\t{eff}+\tfrac{1}{2}) = \langle p^2/(2m) + 
m \Omega_\t{eff}^2 x^2/2 \rangle$,
where $x$ ($p$) is the physical displacement (momentum) of the oscillator at the
trap frequency.  Assuming the displacement and momentum to be zero-mean, their
second moments can be estimated as the integral of their spectral densities.
In principle, two factors complicate this procedure: at lower
frequencies, structural damping renders the displacement variance singular
\cite{Saul90}, while at higher frequencies, feedback back-action precludes a
finite momentum variance \cite{VitTom03}.  In practice, the feedback filter
$\chi_\t{fb}^{-1} \propto \Omega_\t{fb}^2 +\ii \Omega \Gamma_\t{fb}$ is
established around \SIrange{100}{200}{Hz} in an envelope that falls-off at
least as $\Omega^{-2}$ (at frequencies below 10 Hz, the interferometer's length
control loop picks up again), which regulates both these problems.  In this
fashion, within \SIrange{100}{200}{Hz}, the trapped oscillator approximately
satisfies the equipartition principle, and so an effective phonon occupation
can be assigned using the physical displacement spectrum:
\begin{equation*}
    n_\t{eff} \approx \int \frac{S_x[\Omega]}{2x_\t{zp}^2}
        \,\frac{\dd\Omega}{2\pi}.
\end{equation*}
Note that the 100 Hz frequency band in which the oscillator is established is 
much larger than the expected decoherence rate of the trapped oscillator, 
$(n_\t{th}[\Omega_\t{eff}]+n_\t{ba}+n_\t{fb}[\Omega_\t{eff}])\Gamma_0[\Omega_\t{eff}] \approx 2\pi 
\cdot 10\, \t{Hz} $.
We evaluate the integral using the physical displacement spectrum reconstructed
from fits to the observed displacement.
The minimum phonon occupation of the 10 kg oscillator,
corresponding to the purple trace in \cref{fig2}B, is thus inferred to be
$\Nph\pm \dNph$; this is equivalent to an effective mode temperature of
$\SI{\Tfinal}{\nano\kelvin}$. 
This demonstration sets a new record for the 
quantum state purity ($\approx 10\%$ ground state fidelity) for an object of such large mass (see \cref{fig:review}).

The preparation of massive objects progressively nearer their ground state opens the door 
for more sophisticated demonstrations and applications of macroscopic quantum phenomena and quantum metrology.
The most intriguing possibility, however, harnesses the ready susceptibility of kg-scale masses to gravitational
forces; with this work, it becomes possible to prepare them in near-quantum states. This hints
at the tantalizing prospect of studying gravitational decoherence on massive quantum systems.

\bibliographystyle{apsrev4-1}
\bibliography{ref_aligo_cooling}

\noindent \textbf{Funding:} 
This material is based upon work supported by NSF’s LIGO Laboratory which is 
a major facility fully funded by the National Science Foundation.
The authors gratefully acknowledge the support of the United States National Science Foundation (NSF) for the construction and operation of the LIGO Laboratory and Advanced LIGO as well as the Science and Technology Facilities Council (STFC) of the United Kingdom, and the Max-Planck-Society (MPS) for support of the construction of Advanced LIGO. Additional support for Advanced LIGO was provided by the Australian Research Council. The authors acknowledge the LIGO Scientific Collaboration Fellows program for additional support. LIGO was constructed by the California Institute of Technology and Massachusetts Institute of Technology with funding from the National Science Foundation, and operates under cooperative Agreement No. PHY-1764464. Advanced LIGO was built under Award No. PHY-0823459. 
EDH is supported by the MathWorks, Inc. 
This paper carries LIGO Document Number LIGO-P2000525.
\textbf{Competing interests:} The authors declare no competing interests. 
\textbf{Authors contributions:} VS, EDH, and NM conceived this project; CW, EDH, SD, and VS designed
and implemented the modifications to the Advanced LIGO detector that enabled the experiment; 
all authors contributed to the running, diagnostics, and calibration of the detector; CW, EDH, and VS
analysed the data; VS wrote the manuscript with help from EDH, CW, and NM; VS developed
the theoretical models, and supervised the project.
Other LIGO collaboration authors contributed to the design, construction and operation of LIGO, the development and maintenance of data handling, data reduction and data analysis.
All authors meet the journal’s authorship criteria and have reviewed, discussed, and commented on the results and the manuscript.
\textbf{Data and materials availability:} All data needed to evaluate the conclusions in the 
paper are present in the paper or the supplementary materials.

\appendix

\section{Model of measurement and feedback}\label{app:model}

The displacement of the oscillator ($\delta x$) responds to a sum of thermal, back-action, and feedback forces:
\begin{equation}\label{eq:x0}
    \chi_0^{-1} \delta x =  \delta F_\t{th} + \delta F_\t{ba} + F_\t{fb}.
\end{equation}
Here, the susceptibility of the oscillator $\chi_0^{-1}[\Omega] = m(-\Omega^2 +\Omega_0^2 +i \Omega
\Gamma_0[\Omega])$ is well approximated by the test
mass pendulum mode at frequency $\Omega_0\approx 2\pi \cdot \SI{\fzero}{\Hz}$,
which is structurally damped, so that its damping rate is frequency dependent:
$\Gamma_0[\Omega]=(\Omega_0/Q_0)(\Omega_0/\Omega)$, with a quality factor $Q_0
\approx \num{e8}$.

The thermal force ($\delta F_\t{th}$) is characterized by its spectral density,
\begin{equation}
    S_F^\t{th}[\Omega] = 4\hbar \left( n_\t{th}[\Omega] +\tfrac{1}{2}\right)\t{Im}\, \chi_0^{-1}[\Omega], 
\end{equation}
where $n_\t{th}[\Omega] \approx k_B T/(\hbar \Omega) \approx 9\cdot 10^{12}
(\Omega_0/\Omega)$ is the average thermal phonon occupation.  The back-action
force ($\delta F_\t{ba}$), arising from radiation pressure quantum
fluctuations, is characterized by
\begin{equation}
    S_F^\t{ba}[\Omega] = \frac{16\hbar\mathcal{F}}{\lambda c}P_\t{cav} \e^{2r_\text{asqz}},
\end{equation}
where $P_\t{cav}\approx \SI{\Parm}{\kW}$ is the mean arm cavity power at
wavelength $\lambda = \SI{1064}{\nm}$, $\mathcal{F} \approx \num{45}$ is the effective finesse of the signal-recycled arm cavities, and $r_\text{asqz}$ quantifies the increase in
quantum fluctuations of the intracavity optical amplitude due to antisqueezing due to 
the phase-squeezed vacuum injected at the interferometer's dark port; here
$10\log_{10}{e^{2r_\text{asqz}}} = \SI{8\pm1}{dB}$ \cite{hao20}. 
The back-action force
can be quantified in terms of an average phonon occupation $n_\t{ba}$ via
$S_F^\t{ba} \equiv 4\hbar n_\t{ba}\, \t{Im}\, \chi_0^{-1}[\Omega_0]$, which
gives $n_\t{ba}\approx \nba$.

The feedback force $F_\t{fb}$ is based on a linear estimate of the oscillator's position, 
\begin{equation}\label{eq:xe}
    \delta x_\t{est} \equiv G (\delta x + \delta x_\t{imp});
\end{equation}
here, $\delta x_\t{imp}$ is the displacement imprecision (due to sensing
noise), and $G$ is the sensing function of the interferometer.  Such an
estimate is obtained only when the interferometer is stabilized at its linear
operating point, achieved by a feedback loop that forces the test mass (modeled
by the actuation function $A$) based on a filtered (by $D_0$) record of the
error signal $x_\t{est}$.
We create an additional feedback path consisting of a digital filter $D_\t{T}$
in series with $D_0$ to produce the trap, and a parallel path consisting of the
digital filter $D_\t{C}$ to cold-damp the trapped oscillator. The combined
feedback force thus exerted is
\begin{equation}\label{eq:Ffb}
\begin{split}
    F_\t{fb} &= A(D_0D_\t{T}+D_\t{C})\delta x_\t{est} + \delta F_\t{fb} \\
        &\equiv -\chi_\t{fb}^{-1}(\delta x + \delta x_\t{imp}) + \delta F_\t{fb};
\end{split}
\end{equation}
here, $\delta F_\t{fb}$ models extraneous force fluctuations due to the
actuator.  Solving for $\delta x$ between \cref{eq:x0,eq:Ffb} gives the
physical displacement fluctuations,
\begin{equation}\label{eq:xeff}
    \delta x = \chi_\t{eff}\left( \delta F_\t{tot} -\chi_\t{fb}^{-1} \delta x_\t{imp} \right);
\end{equation}
here, $\chi_\t{eff} \equiv (\chi_0^{-1} + \chi_\t{fb}^{-1})^{-1}$ is the effective susceptibility of the oscillator, and
$\delta F_\t{tot} \equiv \delta F_\t{th}+\delta F_\t{ba} +\delta F_\t{fb}$ is the total force noise.

The oscillator can be trapped and cooled by synthesizing an effective susceptibility of the form,
\begin{equation}
    \chi_\t{eff}^{-1} = m(-\Omega^2 +\Omega_\t{eff}^2 +i \Omega \Gamma_\t{eff}).
\end{equation}
We are able to do this by careful design of the effective loop filter $\chi_\t{fb}^{-1}$, which is switched on
in a sequence that both traps the oscillator, and keeps the interferometer unconditionally stable.

\begin{figure}[t!]
    \centering
    \includegraphics[width=0.75\columnwidth]{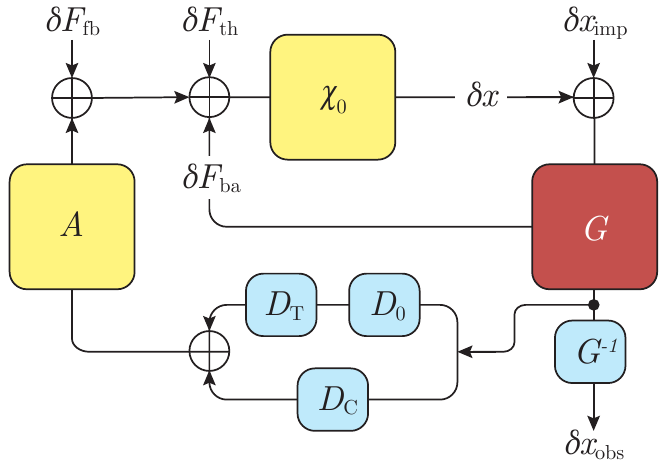}
    \caption{\label{fig:feedback}
    Schematic of the physical system consisting of the intrinsic mechanical response $\chi_0$, the interferometer's
    sensing function $G$, the digital filters $D_\t{0,T,C}$, and the actuation $A$; $G^{-1}$ denotes the
    digital filter used to reconstruct the apparent displacement.
    The feedback is subject to fluctuations arising from actuator force noise $\delta F_\t{fb}$, thermal noise $\delta F_\t{th}$, back-action noise $\delta F_\t{ba}$ and imprecision noise $\delta x_\t{imp}$.
    \label{fig:schematic}}
\end{figure}

What we observe is the apparent displacement $\delta x_\t{obs}$ inferred
using an estimate for the inverse sensing function $G^{-1}$ (that forms part of
LIGO's calibration pipeline \cite{ligo_cal_17}).  That is, $\delta x_\t{obs}
\equiv G^{-1} \delta x_\t{est} \approx \delta x + \delta x_\t{imp}$; amplitude
uncertainty in this estimate is at the $2\%$ level \cite{Sun20}. Using the
known expression for the physical displacement in \cref{eq:xeff}, the apparent
displacement is,
\begin{equation}
    \delta x_\t{obs} = \chi_\t{eff}\left(\delta F_\t{tot} +\chi_0^{-1} \delta x_\t{imp}\right). 
\end{equation}
This model produces the spectrum of the observed signal,
\begin{equation}
\begin{split}
    S_x^\t{obs}[\Omega] = &
    \frac{S_F^\t{tot}[\Omega]/m^2}{(\Omega_\t{eff}^2-\Omega^2)^2 + (\Omega \Gamma_\t{eff})^2} \\
    &+ \frac{(\Omega_0^2-\Omega^2)^2 + (\Omega \Gamma_0[\Omega])^2}
    {(\Omega_\t{eff}^2-\Omega^2)^2 + (\Omega \Gamma_\t{eff})^2}
    S_x^\t{imp}[\Omega],
\end{split}
\end{equation}
that is used to fit the data in \cref{fig2}B in the main text. However the
apparent motion\,---\,since it contains correlations impressed by the feedback
of imprecision\,---\,cannot be directly compared to the spectrum of a physical
oscillator that is damped. 

The spectrum of the physical motion of the oscillator ($\delta x$ in \cref{eq:xeff}), 
\begin{equation}
\begin{split}
    S_x[\Omega] &= \frac{S_F^\t{tot}[\Omega]/m^2}{(\Omega_\t{eff}^2-\Omega^2)^2 + (\Omega \Gamma_\t{eff})^2} \\
    &+ \frac{\Omega_\t{fb}^4 + (\Omega \Gamma_\t{fb})^2}
    {(\Omega_\t{eff}^2-\Omega^2)^2 + (\Omega \Gamma_\t{eff})^2}
    S_x^\t{imp}[\Omega],
\end{split}
\end{equation}
can be directly compared against that of an oscillator trapped at frequency
$\Omega_\t{eff}$, and featuring a damped linewidth $\Gamma_\t{eff}$. Writing
$x=x_\t{zp}(b+b^\dagger)$ for the position of such an oscillator, with
zero-point motion $x_\t{zp}=\sqrt{\hbar/2m\Omega_\t{eff}}$ and creation
operator $b$, presumed to exist in a thermal state, we use the identities,
\begin{align}
    \t{Var}[x] &= 2x_\t{zp}^2 \left( \avg{b^\dagger b}+\tfrac{1}{2} \right) \\
    \t{Var}[x] &= \int S_x[\Omega]\, \frac{\dd\Omega}{2\pi},
\end{align}
to assign an effective phonon number $n_\t{eff}\equiv \avg{b^\dagger b}$:
\begin{equation}
    n_\t{eff}+\tfrac{1}{2} = \int \frac{S_x[\Omega]}{2x_\t{zp}^2}\, \frac{\dd\Omega}{2\pi}.
\end{equation}
Here, the integral is understood to be evaluated in the frequency interval
where the oscillator susceptibility is realized.

When the imprecision noise is white (i.e. $S_x^\t{imp}[\Omega]\approx
S_x^\t{imp}[\Omega_\t{eff}]$), and the frequency-dependence of the
structurally-damped thermal phonon number can be neglected (i.e.
$n_\t{th}[\Omega] \approx n_\t{th}[\Omega_\t{eff}]$)\;---\;both inapplicable to
the current experiment, but useful to develop intuition\;---\;the phonon
occupation can be explicitly evaluated as,
\begin{equation}
\begin{split}
    n_\t{eff}+\frac{1}{2} \approx& \left(n_\t{tot}[\Omega_\t{eff}] 
    + \left(\frac{\Omega_\t{eff}}{\Gamma_0[\Omega_\t{eff}]}\right)^2 n_\t{imp}+ \frac{1}{2}\right)  
    \frac{\Gamma_0[\Omega_\t{eff}]}{\Gamma_\t{eff}[\Omega_\t{eff}]}\\ 
    &\quad + n_\t{imp} \frac{\Gamma_\t{eff}[\Omega_\t{eff}]}{\Gamma_0[\Omega_\t{eff}]}, 
\end{split}
\end{equation}
where the factor in the parentheses in the first line is the total initial
occupation, consisting of the sum of thermal and back-action quanta
($n_\t{tot}=n_\t{th}+n_\t{ba}$), and an additional contribution
$(\Omega_\t{eff}/\Gamma_0)^2 n_\t{imp}$ due to fluctuations in the trap from
feedback of imprecision noise due to the active spring. Here, $n_\t{imp} \equiv
S_x^\t{imp}[\Omega_\t{eff}]/2S_x^\t{zp}[\Omega_\t{eff}]$, is the
phonon-equivalent imprecision ($S_x^\t{zp}[\Omega_\t{eff}] =
8x_\t{zp}^2/\Gamma_0[\Omega_\t{eff}]$ is the peak zero-point spectrum of the
trapped oscillator).

\subsection{Effect of actuator force noise}\label{app:Ffb}

It has been documented that the electrostatic drive (ESD) that is used to actuate the test masses produces excess
force noise that arises from a combination of charging effects and driver voltage noise \cite{Buik19}.
In the context of feedback cooling the test mass, this force (termed $\delta F_\t{fb}$ in \cref{eq:xeff}) acts
as an excess thermal force that heats the trapped oscillator, resulting in additional phonons ($n_\t{fb}$) that 
add to the thermal occupation.
From Ref. \cite{Buik19}, it can be inferred that
\begin{equation*}
    \sqrt{S_\t{F}^\t{fb}}\approx (4\cdot 10^{-18}\, \t{N/\sqrt{Hz}})\, \left(\frac{10\, \t{Hz}}{f} \right).     
\end{equation*}
(By comparison, the typical actuation strength used to keep the interferometer locked is 
$\sim 10^{-6}\, \t{N/\sqrt{Hz}}$ at 10 Hz.)
An extraneous phonon occupation $n_\t{fb,ex}$ can be associated with this force noise via,
$S_F^\t{fb} \approx 4\hbar m \Gamma_0 \Omega_\t{eff} n_\t{fb,ex}$. Assuming
the oscillator is trapped at $\Omega_\t{eff}\approx 2\pi\cdot \feff$ Hz, the equivalent phonon occupation 
from excess ESD noise is, $n_\t{fb,ex} \lesssim 10^{-3}$.

\subsection{Effect of filter delay}

If the feedback is implemented with an overall delay $\tau$\;---\;for example arising from delays in the computation
of the digital filter\;---\;the trapping and cooling filter is modified to
\begin{equation*}
\begin{split}
    \chi_\t{fb}^{-1} e^{i\Omega \tau} &= m(\Omega_\t{fb}^2 + i\Omega \Gamma_\t{fb})e^{i\Omega \tau} \\
    &\approx m [\Omega_\t{fb}^2(1-\tau\Gamma_\t{fb}) + i\Omega(\Gamma_\t{fb}+\tau \Omega_\t{fb}^2)],
\end{split}
\end{equation*}
where in going to the second line, we assume
that the delay is small compared to the characteristic frequency at which it occurs, i.e. $\Omega \tau \ll 1$. 
Thus, even in the absence of active damping (i.e. $\Gamma_\t{fb}=0$), delay in the loop manifests as 
damping $\tau \Omega_\t{fb}^2$. This serves to stabilize the trapped oscillator.

Delay in other parts of the loop manifest as an overall phase factor in the closed-loop gain, 
$\chi_\t{eff} e^{i\Omega \tau}$. Fits to the phase response of the closed-loop gain in the main text resolve
this overall phase shift at the level of $\approx \SI{\tdelay}{\milli\second}$, consistent with expected delays in the loop.

\section{Data analysis}

\subsection{Uncertainty in transfer function fits}\label{app:unc}

The standard deviation in the transfer function estimate $\hat{G}$ derived from signals with coherence $C$ over $N$ averages is given by~\cite{bendat2011random}
\begin{equation}
    \sigma_{\hat{G}} = \sqrt{\frac{(1-C)}{2CN}} \hat{G}.
\end{equation}
Similarly, the coherence estimate $\hat{C}$ has standard deviation
\begin{equation}
    \sigma_{\hat{C}} = \sqrt{\frac{2C}{N}} (1-C).
\end{equation}
Even assuming a worst-case true coherence $C$ within this range, most of the uncertainty in our data arises from the \SI{1}{\second} average duration and corresponding \SI{1}{\hertz} bin width.

We fit to a time-delayed resonator model,
\begin{equation}
    \chi_\text{eff} \sim \frac{1}{\Omega_0^2 - \Omega^2 + \ii \Omega_0 \Omega/Q} \e^{\ii(\phi - \Omega t)},
\end{equation}
and propagate these uncertainties using orthogonal distance regression~\cite{boggs1988odr}.


\end{document}